\documentstyle[12pt]{article}

\def\lromn#1{\uppercase\expandafter{\romannumeral#1}}

\def\blist{\begin{list}{\setlength{\rightmargin}{\leftmargin}}}
\def\elist{\end{list}}
\begin{document}
\begin{flushright}
TU/99/578\\
RCNS-99-06\\
\end{flushright}

\begin{center}
\begin{large}

\bf{
New Kinetic Equation for Pair-annihilating Particles:
Generalization of the Boltzmann Equation
}

\end{large}

\vspace{36pt}

\begin{large}
Sh. Matsumoto and M. Yoshimura

Department of Physics, Tohoku University\\
Sendai 980-8578 Japan\\
\end{large}

\vspace{4cm}

{\bf ABSTRACT}

\end{center}

A convenient form of kinetic equation is derived for pair annihilation
of heavy stable particles relevant to the dark matter problem
in cosmology. 
The kinetic equation thus derived extends the on-shell Boltzmann equation
in a most straightforward way, including the off-shell effect.
A detailed balance equation for the equilibrium abundance is further analyzed.
Perturbative analysis of this equation supports a previous result for the
equilibrium abundance using the thermal field theory, and
gives the temperature power dependence of equilibrium value
at low temperatures. 
Estimate of the relic abundance is possible using this new
equilibrium abundance in the sudden freeze-out approximation.


\newpage
\section{Introduction}

\vspace{0.5cm} 
\hspace*{0.5cm} 
Generalization of the Boltzmann equation is important both from
the point of fundamental physics and of 
application to many areas of physics.
Since its investigation involves quantum dynamics at finite time, 
its general formulation
in a form of practical use is undoubtedly difficult. 
One important classs of physical situations however seems amenable to 
relatively straightforward analysis.
It is a small system immersed in a large thermal environment.

Application of this class of physical situations
includes pair-annihilation of heavy stable
particles in cosmology; the dark matter problem \cite{review on lsp}.
The conventional practice for estimate of the relic abundance
of dark matter particles is first to compute the freeze-out temperature
below which the annihilation effectively ceases due to rapid
cosmological expansion, and then to use the ideal gas form
of number density at that temperature.
A thermally averaged Boltzmann equation is a basis for such computations
\cite{lee-weinberg}.
Clearly, the use of the on-shell Boltzmann equation and 
the ideal gas distribution function should
be questioned, especially at low temperatures such as $T < M/30$ 
($M$ being the mass of the heavy particle) with very small $e^{-M/T}$
like $10^{-\,13}$, which happens to be the relevant
temperature region for the dark matter annihilation.

As in our previous studies 
\cite{my-pair-99},\cite{jmy-decay1},\cite{jmy-decay2},
we take in the present work of
quantum kinetic equation the method of integrated-out
environment in much the same way as some treatment of quantum
Brownian motion \cite{caldeira-leggett}.
This way one can well cope with quantum mechanical behavior at finite
times beyond the infinite time behavior taken care of in 
the S-matrix approach of the Boltzmann equation.
In this paper we report a great and practical simplification over
our previous published result \cite{my-pair-99}. 
We are thus ready for a realistic study of supersymmetric
dark matter problem.

The equilibrium number density thus computed agrees with that
obtained using the thermal field theory in \cite{my-pair-99-2}.
The result of the present work is however more general;
we present the full quantum kinetic equation, not just the
equilibrium result.
The essential ingredient in this formulation is the Hartree
approximation, which makes it possible to use the result
of the solvable model \cite{jmy-decay1}, with an extra assumption
of slow variation of physical quantities.

\newpage
\section{Model of pair annihilation}

\hspace*{0.5cm}
Suppose that
a pair of heavy particles $\varphi $ (boson in this case)
can annihilate into a $\chi $ (boson) pair;
$\varphi \varphi \rightarrow \chi \chi $ with a dimensionless
interaction strength $\lambda $.
A discrete symmetry under $\varphi \rightarrow - \varphi $ is
imposed to forbid the $\varphi $ decay, $\varphi \rightarrow \chi \chi $.
We assume that the lighter particle $\chi $ makes up a thermal 
environment of temperature $T = 1/\beta $ in our unit of
the Boltzmann constant
\( \:
k_{B} = 1
\: \).
Thermalization becomes possible either due to their own self-interaction 
characterized by a dimensionless coupling $\lambda _{\chi }$, or
interaction with other light particles which need not to be specified
for our purpose.
A relativistic field theory model we consider is thus given by
the Lagrangian density
\begin{eqnarray}
&&
{\cal L} = {\cal L}_{\varphi } + {\cal L}_{\chi } +
{\cal L}_{{\rm int}}  \,, 
\\ &&
{\cal L}_{\varphi } + {\cal L}_{\chi } = \frac{1}{2}\,
(\partial _{\mu }\varphi )^{2} - \frac{1}{2}\, M^{2}\,\varphi ^{2}
+ \frac{1}{2}\, (\partial _{\mu }\chi )^{2} - 
\frac{1}{2}\, m^{2}\,\chi ^{2} \,,
\\ &&
{\cal L}_{{\rm int}} = -\,\frac{\lambda }{4}\,\varphi ^{2}\,\chi ^{2}
- \frac{\lambda_{\varphi}}{4!}\,\varphi ^{4}
- \frac{\lambda_{\chi}}{4!}\,\chi ^{4} + \delta {\cal L}
\,.
\end{eqnarray}
The last two terms ($\propto \lambda _{\varphi }$ and $\lambda _{\chi }$)
in the interaction are introduced both for consistency
of renormalization and for thermalization of $\chi $ particles, 
hence the parameters are taken to satisfy
\( \:
|\lambda_{\varphi}|  \ll   \lambda^2 \ll 1
\,, \hspace{0.2cm} \; |\lambda_{\chi}| < 1 \,,
\: \)
but $|\lambda _{\chi }| \gg |\lambda |$.
The renormalization counter term $\delta {\cal L}$ is given by
\begin{equation}
        \frac{\delta Z_\varphi}{2}(\partial_{\mu}\varphi)^2
        - \frac{\delta M^2}{2}\varphi^2
        + \frac{\delta Z_\chi}{2}(\partial_{\mu}\chi)^2
        - \frac{\delta m^2}{2}\chi^2
        - \frac{\delta \lambda}{4}\varphi^2\chi^2
        - \frac{\delta \lambda_{\varphi}}{4!}\varphi^4
        - \frac{\delta \lambda_{\chi}}{4!}\chi^4
\,.
\end{equation}

We focuss on a particular
dynamical degree of freedom, the heavy $\varphi $ field, 
and integrates out the environment part, the $\chi $ field,
altogether. 
In the influence functional method \cite{feynman-vernon} 
this integration is carried out for the squared amplitude,
namely for the probability function, by using the path integral technique.
This way one has to deal with the conjugate field variable $\varphi '$
along with $\varphi $, since the complex conjugated quantity is
multiplied in the probability.
The influence functional ${\cal F}$ is thus defined by 
\begin{eqnarray}
&& 
\int\,{\cal D}\chi  \,\int\,{\cal D}\chi '\,\exp 
\left[\, i \int\,dx\,
\left( \,{\cal L}_{\chi }(x) - {\cal L}_{\chi '}(x)
\right.
\right.
\nonumber \\ && \hspace*{0.5cm} \left. \left.
+\, {\cal L}_{{\rm int}}(\varphi(x) \,, \chi (x))
- {\cal L}_{{\rm int}}(\varphi'(x) \,, \chi' (x))
\,\right) \rule{0mm}{2.5ex} \right] \,.
\end{eqnarray}

We convolute with the influence functional 
the initial and the final density matrix of the $\chi  $ system.
For the initial state it is assumed that the entire system
is described by an uncorrelated product of the system and the
environment density matrix,
\begin{equation}
\rho _{i} = 
\rho _{i}^{(\varphi )} \times \rho _{i}^{(\chi )} \,, \hspace{0.5cm} 
\rho _{i}^{(\chi )} = 
\rho _{\beta }^{(\chi)} = e^{-\,\beta H_{0}(\chi )}/{\rm tr}\,
e^{-\,\beta H_{0}(\chi )} \,.
\label{initial density matrix} 
\end{equation}
Here $\rho _{\beta }^{(\chi)}$ is the density matrix for a thermal
environment.
The density matrix $\rho _{i}^{(\varphi )}$
for the $\varphi $ system is arbitrary.
The choice of the initial density matrix $\rho _{i}$ is not crucial
and the final result is expected to be insensitive to this choice
provided that it does not commute with the total hamiltonian,
$[\rho _{i} \,, H] \neq 0$.
We need this condition to guarantee a departure from complete
thermal equilibrium.
At a final time $t_{f}$ the $\chi $ integration
is performed taking the condition of non-observation for the environment,
\begin{eqnarray}
\int\,d\chi _{f}\,\int\,d\chi _{f}'\,\delta (\chi _{f} - \chi_{f}')\,
(\cdots ) \,, 
\end{eqnarray}
with the understanding that the environment is totally unspecified
at the time $t_{f}$.

The result of $\chi $ integration is given by a series of Gaussian integral
if we expand in powers of $\lambda _{\chi }$ of the 
$\lambda _{\chi }\,\chi ^{4}$ interaction,
and the influence functional to order $\lambda ^{2}$  is of the form,
\begin{eqnarray}
&& \hspace*{1cm}
{\cal F}_{4}[\varphi \,, \varphi '] =
\nonumber \\ && \hspace*{-1cm}
\exp [\,-\,\frac{1}{4}\,\int_{x_{0} > y_{0}}\,dx\,dy\,
\left( \,\xi _{2}(x)\,\alpha _{R}(x\,,  y)\,\xi _{2}(y) + i\,
\xi _{2}(x)\,\alpha _{I}(x \,,  y)\,X_{2}(y)\,\right)\,] \,, 
\label{original influence functional} 
\\ &&
\hspace*{1cm} 
X_{2}(x) \equiv  \varphi ^{2}(x) + \varphi '\,^{2}(x) 
\,, \hspace{0.5cm} 
\xi _{2}(x) \equiv \varphi ^{2}(x) - \varphi '\,^{2}(x) \,, 
\\ && 
\hspace*{1cm} 
\alpha (x \,, y) = \alpha _{R}(x \,, y) + i\alpha _{I}(x \,, y)
\nonumber \\ &&
=\, \lambda ^{2}\,\left( \,
{\rm tr}\;\left( T[\chi^{2}(x) \chi^{2}(y)\,
\rho _{\beta }^{(\chi )}]\right) 
- (\,{\rm tr}\; \chi ^{2}\rho _{\beta }^{(\chi)}\,)^{2}
\,\right) \,,
\label{2-point kernel} 
\end{eqnarray}
if one replaces the mass term in the original Lagrangian by a
$O[\lambda ]$ temperature dependent mass, 
$M^{2}(T) = M^{2} + \frac{\lambda }{24}\,T^{2}$ valid for $m \ll T$.
Note the presence of the time ordering, $x_{0} > y_{0}$, in the above
formula.
The kernel function $\alpha (x\,,  y)$ satisfies
the time translation invariance, thus may be written
as $\alpha (x - y)$. 
An explicit form of the
kernel function $\alpha (x)$ or its Fourier transform is given later
in eqs.(\ref{fv kernel for 2-body}) and (\ref{two-body spectral}).
Higher order terms in $\lambda ^{2}$ are actually present in the 
exponent of the influence functional. These contribute either to
many-body processes we are not interested in, or to negligible higher order
terms to our process.

The convolution with the heavy $\varphi $ system  gives the
reduced density matrix at time $t_{f}$;
\begin{eqnarray}
&& \hspace*{-1cm}
\rho ^{(R)}(\varphi _{f} \,, \varphi _{f}') =
\,\int\,d\varphi _{i}\,\int\,d\varphi_{i}'
\int\,{\cal D}\varphi \,\int\,{\cal D}\varphi '
\,e^{iS(\varphi ) - iS(\varphi ')}\,
{\cal F}_{4}[\varphi \,, \varphi ']\,\rho _{i}^{(\varphi )}
(\varphi _{i} \,, \varphi _{i}')
 \,, \nonumber \\ &&
\end{eqnarray}
from which one can deduce physical quantities for the $\varphi $ system.
Here $S(\varphi )$ is the action for the $\varphi $ system obtained
from the basic Lagrangian.

The correlator is defined using the reduced density matrix;
for $x_{0} > y_{0}$
\begin{eqnarray}
&& 
\langle \,\varphi (x) \varphi (y)\, \rangle =
\int\,d\varphi (x)\,\int\,d\varphi'(x)
\,\int\,d\varphi (y)\,\int\,d\varphi '(y)\,\delta \left( \varphi (x) -
\varphi' (x)\right)
\nonumber \\ && \hspace*{-1cm} 
\cdot \int\,{\cal D}\varphi \,\int\,{\cal D}\varphi '
\,e^{iS(\varphi ) - iS(\varphi ')}\,
{\cal F}_{4}[\varphi \,, \varphi ']\,\varphi (x)\varphi (y)
\,\rho ^{(R)}(\varphi (y) \,, \varphi' (y))
\,. 
\end{eqnarray}
The path to be taken here is those defined on the time interval,
$x_{0} > t > y_{0}$.
A nice feature of this correlator formula is that the initial memory
effect appears compactly via the reduced density matrix $\rho ^{(R)}$.
With the Heisenberg evolution $\varphi (x) = e^{iHt}\varphi (\vec{x} \,, 0)
e^{-iHt}$, this correlator is equal to
\begin{equation}
{\rm tr}\; \rho _{i}\,\varphi (x)\varphi (y) \,, 
\end{equation}
where $\rho _{i}$ is the total initial density matrix 
(\ref{initial density matrix}).
Whenever this density matrix does not commute with the total
hamiltonian,
$[\rho _{i} \,, H] \neq 0$,
this correlator does not respect the time-translation invariance,
\begin{equation}
\langle \varphi (\vec{x} \,, t_{1})\varphi (\vec{y} \,, t_{2}) \rangle
\neq 
\langle \varphi (\vec{x} \,, t_{1} - t_{2})\varphi (\vec{y} \,, 0) \rangle
\,,
\end{equation}
unlike the correlator in complete thermal equilibrium.

The model thus specified is difficult to solve due to the appearance of
the quartic term of $\varphi $ in the influence functional 
(\ref{original influence functional}). 
The situation is however simplified when one
considers a mean field approximation. 
In the mean field or the Hartree approximation one replaces
an even number of field operators by a pair of two operators times
the rest of several averaged two-body correlators.
This approximation is good if one can ignore a higher order correlation
than that of two-body. 
The Hartree model is expected to work well in a dilute system
characterized by a low occupation number for each mode. 

The Hartree approximation we introduce is then
a Gaussian truncation to
the influence functional; we replace the original one by properly defining
new real kernel functions $\beta_{i}(x\,, y)$ in the quadratic form;
\begin{eqnarray}
&& \hspace*{-1cm}
{\cal F}_{2}[\varphi \,, \varphi '] =
\exp [\,-\,\int_{x_{0} > y_{0}}\,dx\,dy\,
\left( \,\xi (x)\,\beta  _{R}(x\,, y)\,\xi (y) + i\,
\xi (x)\,\beta  _{I}(x\,, y)\,X(y)\,\right)\,] \,, 
\label{influence functional in r-model} 
\\ &&
\hspace*{1cm} 
X(x) \equiv \varphi (x) + \varphi '(x) \,, \hspace{0.3cm}
\xi (x) \equiv \varphi (x) - \varphi '(x) \,.
\end{eqnarray}
When the time-translation invariance holds for $\beta _{i}$,
one can work out its consequences to all order of $\lambda $
for a given kernel $\beta_{i}(x \,, y)$,
since it becomes a solvable Gaussian model.

For derivation of the self-consistency relation 
introduce the full propagator by
\begin{eqnarray}
&&
G(x \,, y) = 
i\,\langle \varphi (x)\varphi (y) \rangle_{{\cal F} = {\cal F}_{4}} \,,
\end{eqnarray}
using the influence functional (\ref{original influence functional}).
We compute correlators,
\begin{equation}
\langle X(x)\xi (y) \rangle \,, \langle X(x)X(y) \rangle
\,, \langle \xi (x) \xi (y)\rangle \,,
\end{equation}
in two theories, using the two
different forms of the influence functional,
${\cal F}_{4}$ and the truncated one ${\cal F}_{2}$,
and identify two results. One then has ($\beta = \beta _{R} + i\beta _{I}$);
\begin{eqnarray}
&&
\beta (x_{1}\,, x_{2}) = -\,i\,\alpha (x_{1} - x_{2})\,G(x_{1}\,, x_{2})
\,.
\label{relation between 2 kernels} 
\end{eqnarray}

It is important to note again that there is no translational
invariance with respect to time variables $t_{1} \,, t_{2}$, 
since the system
is not in complete thermal equilibrium.
Thus, one should distiguish two time variables for the two-point
correlator, the relative time $\tau = t_{1} - t_{2}$ and
the the central time $t = (t_{1} + t_{2})/2$.
A fundamental strategy in the present work is
to note the slow change of physical quantities for the central time 
and exploit it in the analysis of the above relation 
(\ref{relation between 2 kernels}).
One thus solves short time dynamics of the relative time
assuming a constant central time.
The result of solvable Gaussian model \cite{jmy-decay1},
\cite{jmy-decay2} may then be used provided that
the kernel function $\beta $ is given.
Since the correlator thus derived is a functional of an yet unknown
$\beta $, one arrives at the self-consistency equation
for this $\beta $, which contains the constant central time.
This constant central time is then allowed to slowly vary in the
spirit of adiabatic approximation.
We shall explain this procedure in detail in due course.

\vspace{1cm}
\section{
Correlation function in Hartree approximation
}

\hspace*{0.5cm}
It is convenient to define the Fourier transform with respect to
the relative variable,
\begin{equation}
      \sigma(k,t) = \int\, d^4(x - y)\, \langle \varphi(x)\,\varphi(y) \rangle
      e^{ik\cdot (x - y)}
\,.
\end{equation}
Since spatial homogeneity holds, 
this quantity $\sigma $ may depend on the central coordinate
only via $t = (x_{0} + y_{0})/2$.

Suppose that the central time dependence is slow, and we may
take this as a constant.
The correlation function can then be calculated using the technique of the
generating functional for the influence functional, as
explained in ref.\cite{my-pair-99}.
In the Gaussian model or in the Hartree model of pair annihilation each
Fourier component of correlators is given in terms of a basic
function $g(\vec{k} \,, t)$ and the density matrix of the
initial state.
Suppressing all Fourier momenta and
denoting $q(t)$ for Fourier component of field and $p(t)$
for its time derivatives, one has, for instance, for the $q-q$ correlator
\begin{eqnarray}
&& 
\langle q(t_{1}) q(t_{2})\rangle =
- \frac{i}{2}\,g(t_{1} - t_{2}) 
+ \int_{0}^{t_{1}}\,dt \,\int_{0}^{t_{2}}\,ds\,
g(t_{1} - t)\beta _{R}(t - s)g(t_{2} - s) 
\nonumber \\ && \hspace*{0.5cm} 
+\, g(t_{1})g(t_{2})\,
\overline{p_{i}^{2}}
+ \dot{g}(t_{1})\dot{g}(t_{2})\,
\overline{q_{i}^{2}} 
+ \frac{1}{2}\,(\,\dot{g}(t_{1})g(t_{2}) + g(t_{1})\dot{g}(t_{2})\,)
\,\overline{\{\,p_{i}\,, q_{i}\,\}}
 \,.
\label{qq correlator at different t} 
\end{eqnarray}
The initial density matrix elements are needed only to sepcify
three expectation values, 
$\overline{p_{i}^{2}} \,, \overline{q_{i}^{2}}$ 
and $\overline{\{\,p_{i}\,, q_{i}\,\}}$.

The basic function $g(t)$
is related to the spectral weight $r(\omega)$, and ultimately
to the Fourier transform of the correlator.
Fourier transforms, $r$ and $r_{\chi }$, 
with respect to the relative coordinate are introduced as
\begin{eqnarray}
&&
\beta (x \,, y) \equiv \int\,\frac{d^{4}k}{(2\pi )^{3}}\,
r(k \,, \frac{x_{0} + y_{0}}{2})\,e^{-ik\cdot (x - y)} \,,
\\ &&
\alpha (x - y) = \int\,\frac{d^{4}k}{(2\pi )^{3}}\,\frac{2}{1 - e^{-\beta 
k_{0}}}\,r_{\chi }(k)\,e^{-ik\cdot (x - y)} \,.
\label{fv kernel for 2-body} 
\end{eqnarray}
One obtains from the self-consistency relation
\begin{eqnarray}
      r(k,t) = 2 \lambda^2 \,\int^{\infty}_{-\infty}\,
      \frac{d^4k'}{(2\pi)^4}\,\frac{r_\chi(k - k')}
      {1 - e^{-\beta(k_0 - k'_0)}}\sigma(k',t)
      \,.
\label{self-consistency in fourier} 
\end{eqnarray}
The important relation between $r(k \,, t)$ and 
$\sigma (k\,, t)$, eq.(\ref{self-consistency in fourier}), is a result of
eq.(\ref{relation between 2 kernels}).
The 3-momentum dependence is always even, being function of
$|\vec{k}| = k$. Thus, $r_{\chi }$ is a function of two variables,
$k_{0} = \omega$ and $k$, hence can be written as $r_{\chi }(\omega  \,, k)$.
It is calculable from Fig.1 \cite{jmy-decay1} and is odd in $\omega $;
for $\omega  > 0$
\begin{eqnarray}
&& \hspace*{-1.5cm}
r_{\chi }(\omega  \,, k) = 
\frac{\lambda ^{2}}{16\pi ^{2}}\,\left( \,
\sqrt{1 - \frac{4m^{2}}{\omega ^{2} - k^{2}}}\,\theta (\omega 
- \sqrt{k^{2} + 4m^{2}}) + \frac{2}{\beta k}\,
\ln \frac{1 - e^{-\beta \omega _{+}}}{1 - e^{-\beta |\omega _{-}|}}\,\right)
\,, \label{two-body spectral} 
\\ && \hspace*{1cm} 
\omega _{\pm } = \frac{\omega }{2} \pm \frac{k}{2}\,
\sqrt{1 - \frac{4m^{2}}{\omega ^{2} - k^{2}}} \,.
\end{eqnarray}

The basic function $g$ that appears in the correlator is then given by
\begin{eqnarray}
&&
      g_t(\vec{k}\,, \tau ) = i\,\int^\infty_{-\infty}\, d\omega\,
      H(\omega,\vec{k},t)\,e^{-i\omega \tau }
 \,,
 \label{fourier of g} 
\\ &&
      H(k\,, t)
      = \frac{r_-(k\,, t)}
      {(k^2 - M^2(T\,, t) - \Pi(k\,, t))^2
        + (\pi r_-(k\,, t))^2}
\,,
\\ &&
\Pi (k\,, t) = {\cal P}\,\int_{-\infty }^{\infty }\,d\omega '\,
\frac{r_{-}(\omega ' ,\vec{k}\,, t)}{\omega ' - k_{0}} \,, 
\label{proper self-energy} 
\end{eqnarray}
where even and odd parts of the spectral weight are defined by
\begin{equation}
r_{\pm  }(k\,, t) = r(k\,,t) \pm r(-k \,, t) \,.
\end{equation}
The time dependence of the mass shift $M^{2}(T\,, t)$ is weak and
its asymptotic equilibrium value $M^{2}(T)$ to $O[\lambda ^{2}]$
is given later by eq.(\ref{mass shift}).
With the help of
\begin{equation}
      h_t(\omega,\vec{k}\,,\tau ) = \int^{\tau }_0\, ds\,
      g_t(\vec{k}, s )e^{-i\omega s}
      = \int^\infty_{-\infty}\,d\omega'\,
      \frac{H(\omega',\vec{k}\,,t)}{\omega - \omega'}
      (e^{i\omega' \tau } - e^{i\omega \tau })
\,, 
\end{equation}
the correlation function is calculated as
\begin{eqnarray}
  &~&
  \langle \varphi(x)\,\varphi(y) \rangle
  \nonumber \\
  &=&
  \int\, \frac{d^3k}{(2\pi)^3}\,
  \left[
    -\frac{i}{2}g_t(\vec{k},x^0 - y^0)
    +
    \int^{\infty}_{-\infty} d\omega\,\frac{r_+(\omega,\vec{k},t)}{2}
    h^*_t(\omega,\vec{k},x^0)h_t(\omega,\vec{k},y^0)
    \right.
    \nonumber \\
    &~&
    +~
    g_t(\vec{k},x^0)g_t(\vec{k},y^0)
    \overline{\dot{q}_{\vec{k}}^2}
    + 
    \dot{g}_t(\vec{k},x^0)\dot{g}_t(\vec{k},y^0)
    \overline{q_{\vec{k}^2}}
    \nonumber \\
    &~&
    \left.
    +
    \frac{1}{2}
    \left\{
      \dot{g}_t(\vec{k},x^0)g_t(\vec{k},y^0)
      + g_t(\vec{k},x^0)\dot{g}_t(\vec{k},y^0)
    \right\}
    \overline{
    \dot{q}_{\vec{k}}q_{-\vec{k}} + q_{\vec{k}}\dot{q}_{-\vec{k}}}
        \rule {0mm}{4ex}
    \right]e^{i\vec{k}\cdot(\vec{x} - \vec{y})}
\,.    \label{correlator form} 
\end{eqnarray}
Note that $t = (x_{0} + y_{0})/2$.
The integrand of 3-momentum integral of this formula is divided into
the antisymmetric part (1st term) and the symmetric part (the rest)
under the interchange $x_{0} \leftrightarrow y_{0}$.
On the other hand, the antisymmetic part of the correlation function
is related to the spectral weight $H$ as seen from eq.(\ref{fourier of g}), 
thus the antisymmetric part of the correlator is given by
\begin{equation}
\sigma _{-}(k \,, t) = 2\pi \,H(k\,, t) \,.
\end{equation}

The equation thus derived (\ref{correlator form})
can be regarded as a self-consistency equation
for $\langle \varphi(x)\,\varphi(y) \rangle$, or equivalently
for its Fourier transform $\sigma (k \,, t)$.
Its derivation rests with the Hartree self-consistency and
the slow variation approximation on the central time.
In the following sections we derive more convenient form of
differential equations that gives an equivalent result.

\vspace{1cm}
\section{Detailed balance and equilibrium abundance}

\hspace*{0.5cm} 
One takes the infinite central time limit of the self-consistency equation
in the preceeding section to derive detailed balance relation,
from which one can analyze the equilibrium abundance.

First, one has for the correlator
\begin{eqnarray}
  \langle \varphi(x)\,\varphi(y) \rangle
  &\longrightarrow&
  \langle \varphi (x)\varphi (y) \rangle_{\infty } = 
  \frac{1}{2}\,\int\,
  \frac{d^4k}{(2\pi)^3}\,
  H(k,\infty)e^{-ik\cdot (x - y)}
  \nonumber \\
  &+&
  \frac{1}{2}\,\int\,
  \frac{d^4k}{(2\pi)^3}\,
  \frac{r_+(k,\infty)}{r_-(k,\infty)}
  H(k,\infty)
  e^{-ik\cdot (x - y)} \,,
\end{eqnarray}
in the infinite central time limit, $(x_{0} + y_{0})/2 \rightarrow \infty $.
All initial memory effects drop out in this limit, as expected.
Its Fourier transform gives odd and even parts,
\begin{eqnarray}
      \sigma_-(k,\infty) = 2\pi H(k,\infty)
\,, \hspace{0.5cm} 
      \sigma_+(k,\infty) =
      \frac{2\pi r_+(k,\infty)}{r_-(k,\infty)}H(k,\infty)
\,,
  \label{even and odd of correlator} 
\end{eqnarray}
since the spectral function $H(k,\infty)$ is odd in $k_{0}$.

We further define a quantity,
\begin{equation}
  \tau(k\,,\infty ) = \frac{\sigma(-k,\infty )}{\sigma_-(k,\infty )} \,.
  \label{def of tau}
\end{equation}
One set of the consistency equation derived by combining 
eqs.(\ref{even and odd of correlator}) and (\ref{def of tau})
\begin{equation}
    r(k,\infty)\tau(k,\infty) - r(-k,\infty)(1 + \tau(k,\infty)) = 0
\,,
\end{equation}
is then written as
\begin{eqnarray}
  \label{detailed balance3}
  0&=&
  16\lambda ^2\int^\infty_0 \frac{dp'_0}{2\pi} \int \frac{d^3p'}{(2\pi)^3}
  \int \frac{d^3k_1}{(2\pi)^32\omega_{k_1}}
  \int \frac{d^3k_2}{(2\pi)^32\omega_{k_2}}\,
  (2\pi)^3\sigma_-(p',\infty)
  \nonumber \\
  &~&
  \left\{ \,
    \delta^{(4)}(p + p' - k_1 - k_2)
    \left[
      \tau \tau'(1 + f_1)(1 + f_2) - (1 + \tau )(1 + \tau')f_1f_2
    \right]
  \right.
  \nonumber \\
  &~& +
  2\delta^{(4)}(p + p' + k_1 - k_2)
  \left[
    \tau \tau'f_1(1 + f_2) - (1 + \tau )(1 + \tau')(1 + f_1)f_2
  \right]
  \nonumber \\
  &~& + 
  \delta^{(4)}(p + p' + k_1 + k_2)
  \left[
    \tau \tau'f_1f_2 - (1 + \tau )(1 + \tau')(1 + f_1)(1 + f_2)
  \right]
  \nonumber \\
  &~& +
  \delta^{(4)}(p - p' - k_1 - k_2)
  \left[
    \tau (1 + \tau')(1 + f_1)(1 +f_2) - (1 + \tau )\tau'f_1f_2
  \right]
  \nonumber \\
  &~&  +
  2\delta^{(4)}(p - p' + k_1 - k_2)
  \left[
    \tau (1 + \tau')f_1(1 + f_2) - (1 + \tau )\tau'(1 + f_1)f_2
  \right]
  \nonumber \\
  &~&
  \left. +
    \delta^{(4)}(p - p' + k_1 + k_2)
    \left[
      \tau (1 + \tau')f_1f_2 - (1 + \tau )\tau'(1 + f_1)(1 + f_2)
    \right]\,
  \right\} \,.
  \label{detailed balance eq} 
\end{eqnarray}
Here
\( \:
\tau  = \tau(p\,, \infty ) \,, \tau' = \tau(p'\,, \infty ) \,, 
\: \)
while
\begin{eqnarray}
&&
\omega _{k} = \sqrt{\vec{k}\,^{2} + m^{2}} 
 \,,
\\ &&
f_1 = f_{{\rm th}}(k_1) \,, \hspace{0.5cm} 
f_2 = f_{{\rm th}}(k_2) \,, \hspace{0.5cm} 
f_{{\rm th}}(k) = \frac{1}{e^{\omega _{k}/T} - 1} \,.
\end{eqnarray}
Note that the momentum $p$ is not integrated in this equation, thus
eq.(\ref{detailed balance eq}) is an integral equation for 
the unknown function $\tau (p\,, \infty )$.

There is an obvious solution for this set of equations:
\begin{equation}
  \tau(p\,,\infty) = \frac{1}{e^{\beta p_0} - 1} \,,
\end{equation}
due to the presence of energy conservation in each associated process
of eq.(\ref{detailed balance eq}).
The 3-momentum dependence in $\tau (p \,, \infty )$ is missing.
We further advance our analysis assuming that this is the unique set of
solutions to the above equation.
Putting these into the other set of consistency equation at
infinite time, one obtains a closed form of self-consistency
equation for $r_{-}$;
\begin{eqnarray}
  r_-(k,\infty) &=& 2\,\lambda^2\int^\infty_{-\infty} \frac{d^4k'}{(2\pi)^3}
  \, H(k' \,, \infty)\,
  r_\chi(k + k')\frac{e^{\beta{k_0}} - 1}
  {(e^{\beta(k_0 + {k'}_0)} - 1)(1 - e^{-\beta{k'}_0})} \,,
  \nonumber \\ \\
  H(k \,, \infty) &=&
  \frac{r_-(k,\infty)}
  {(k^2 - M^2(T) - \Pi(k,\infty))^2 + (\pi r_-(k,\infty))^2}
   \,.
   \label{overlap formula} 
\end{eqnarray}
The quantity $\Pi (k \,, \infty)$ is given by a principal value
integral of $r_{-}(k \,, \infty)$ like (\ref{proper self-energy}).
The other independent quantity $r_{+}$ is determined by
\begin{equation}
r_{+}(k\,, \infty ) = \coth \frac{\beta k_{0}}{2}\,
r_{-}(k\,, \infty ) \,,
\end{equation}
hence giving the correlator
\begin{equation}
\langle \varphi(x) \varphi(y) \rangle_{\infty } =
\int\,\frac{d^4 k}{(2\pi)^3}\,\frac{1}{1 - e^{-\,\beta k_{0}}}\,
H(k \,, \infty) \,e^{-i\,k\cdot(x - y)} \,.
\label{asymp thermal correlator} 
\end{equation}
This way one can completely determine $r(k\,, \infty )$, hence
$\sigma (k\,, \infty )$.

One may first note the trivial case of application of 
eq.(\ref{asymp thermal correlator}),
taking the zero coupling limit $r = 0$.
In this case one should understand the quantity $H$
as 
\begin{equation}
H(k \,, \infty) = \delta (k_{0}^{2} - E_{k}^{2})\,
\epsilon (k_{0}) \,, 
\end{equation}
with 
\begin{equation}
E_{k} = \sqrt{\vec{k}^2 + M^2} \,, 
\end{equation}
thus gets the free field correlator at finite temperature $T$.

Next, perturbative analysis of the detailed balance relation shows
that to $O[\lambda ^{2}]$ 
\begin{eqnarray}
  r_-(k\,,\infty)
  &=&
   \,2\lambda^2\,\int\,\frac{d^{3}k'}{(2\pi )^{3}\,E_{k'}}
  \left[\,
    r_\chi(k_0 + E_k,\vec{k} + \vec{k}')\,
    \frac{e^{\beta k_{0}} - 1}
    {(e^{\beta(k_{0}  + E_{k'})} - 1)(1 - e^{-\beta E_{k'}})}
  \right.
  \nonumber \\
  &~& 
  + \,
  \left.
    r_\chi(k_0 - E_k,\vec{k} - \vec{k}')\,
    \frac{e^{\beta k_{0}} - 1}
    {(e^{\beta(k_{0} - E_{k'})} - 1)(e^{\beta E_{k'}} - 1)}
\,  \right] \,.
\label{perturbative spectral} 
\end{eqnarray}
The spectral function $r_{-}$ given by (\ref{perturbative spectral})
may be written in a suggestive form by using an expression
for $r_{\chi }$ of \cite{weldon}. It is equivalent to
\begin{eqnarray}
&&
  r_-(p\,,\infty) = \frac{\lambda ^{2}}{2}\,\int\,
  d\Pi_{p'}\,\int\,d\Pi _{k}\,\int\,d\Pi _{k'}\,
  (2\pi )^{3}\,\delta ^{(3)}(\vec{p} + \vec{p}\,' + \vec{k} + \vec{k}\,')
  \nonumber 
\\ &&
\{\,
\delta (p_{0} + E_{p'} - \omega _{k} - \omega _{k'})\,
[\tilde{f}_{p'}(1 + f_{k})(1 + f_{k'}) - (1 + \tilde{f}_{p'})f_{k}f_{k'}]
\nonumber 
\\ &&
+\,
2\,\delta (p_{0} + E_{p'} - \omega _{k} + \omega _{k'})\,
[\tilde{f}_{p'}(1 + f_{k})f_{k'} - (1 + \tilde{f}_{p'})f_{k}(1 + f_{k'})]
\nonumber 
\\ &&
+\,
\delta (p_{0} + E_{p'} + \omega _{k} + \omega _{k'})\,
[\tilde{f}_{p'}f_{k}f_{k'} - (1 + \tilde{f}_{p'})(1 + f_{k})(1 + f_{k'})]
\nonumber 
\\ &&
+\,
\delta (p_{0} - E_{p'} - \omega _{k} - \omega _{k'})\,
[(1 + \tilde{f}_{p'})(1 + f_{k})(1 + f_{k'}) - \tilde{f}_{p'}f_{k}f_{k'}]
\nonumber 
\\ &&
+\,
\delta (p_{0} - E_{p'} + \omega _{k} + \omega _{k'})\,
[(1 + \tilde{f}_{p'})f_{k}f_{k'} - \tilde{f}_{p'}(1 + f_{k})(1 + f_{k'})]
\nonumber 
\\ &&
+\,
2\,\delta (p_{0} - E_{p'} + \omega _{k} - \omega _{k'})\,
[(1 + \tilde{f}_{p'})f_{k}(1 + f_{k'}) - \tilde{f}_{p'}(1 + f_{k})f_{k'}]
\,\} \,.
\label{r_ using f} 
\end{eqnarray}
The function $\tilde{f}_{p}$ was introduced here,
\begin{equation}
\tilde{f}_{p} = \frac{1}{e^{E_{p}/T} - 1} \,, 
\end{equation}
thermal $\chi $ particle distribution function 
being written as $f_{k} \,, f_{k'}$.
The phase space factor here is
\begin{equation}
d\Pi _{p} = \frac{d^{3}p}{(2\pi )^{3}\,2E_{p}} \,, \hspace{0.5cm} 
d\Pi _{k} = \frac{d^{3}k}{(2\pi )^{3}\,2\omega _{k}} \,.
\end{equation}
An important point to note is that this spectral function $r_{-}$
is defined even off
the mass shell $p_{0} \neq \sqrt{\vec{p}\,^{2} + M^{2}}$.
Since the negative $p_{0}$ region gives
$r_{-}(-p_{0}, \vec{p} \,, \infty ) = -\,r_{-}(p_{0},
\vec{p} \,, \infty )$, one should be careful in associating indivisual
terms of eq.(\ref{r_ using f}) with physical processes.

The correlator to $O[\lambda ^{2}]$ is obtained using this $r_{-}$
in eq.(\ref{asymp thermal correlator}) along with (\ref{overlap formula}).
This result can be compared to a calculation using 
the thermal field theory in \cite{my-pair-99-2}.
The imaginary-time formalism gives the correlator to
$O[\lambda ^{2}]$,
\begin{eqnarray}
  &~&
  \langle \varphi(\tau_1,\vec{x}_1)\,\varphi(\tau_2,\vec{x}_2) \rangle_\beta
  \nonumber \\
  &=&
  \Delta_\varphi(x_1 - x_2)
  -\frac{\lambda}{2}\int^\beta_0 d^4y\,
  \Delta_\chi(0)\Delta_\varphi(y - x_1)\Delta_\varphi(y - x_2)
  \nonumber \\
  &+&
  \frac{\lambda^2}{2}\int^\beta_0 d^4y_1~d^4y_2\,
  \Delta_\varphi(y_1 - y_2)\Delta_\chi(y_1 - y_2)\Delta_\chi(y_1 - y_2)
  \Delta_\varphi(y_1 - x_1)\Delta_\varphi(y_2 - x_2)
  \nonumber \\
  &+&
  \frac{\lambda^2}{4}\int^\beta_0 d^4y_1~d^4y_2\,
  \Delta_\varphi(0)\Delta_\chi(y_1 - y_2)\Delta_\chi(y_1 - y_2)
  \Delta_\varphi(y_1 - x_1)\Delta_\varphi(y_1 - x_2)
  \nonumber \\
  &+&
  \frac{\lambda^2}{4}\int^\beta_0 d^4y_1~d^4y_2\,
  \Delta_\chi(0)\Delta_\chi(0)\Delta_\varphi(y_1 - y_2)
  \Delta_\varphi(y_1 - x_1)\Delta_\varphi(y_2 - x_2)
  \label{thermal correlator} 
\\
  &\simeq&
  \int \frac{d^3p}{(2\pi)^3}\,T\sum_{n}
  e^{-\omega_{n}(\tau _1 - \tau _2)}e^{i\vec{p}\cdot (\vec{x}_1 - \vec{x}_2)}
  F(\omega_{n},\vec{p}) \,.
\end{eqnarray}
where the variable $\tau $ is the Euclidean time defined in the range of
$0 \sim \beta $, and $\omega_{n} = 2\pi in/\beta \, (n = 0\,, 
\pm 1 \,, \pm 2 \cdots )$.
The quantities $\Delta _{\varphi }$ and $\Delta _{\chi }$ are
the thermal propagators of the $\varphi $ and $\chi $ particles,
respectively.
Each term in eq.(\ref{thermal correlator}) corresponds to respective
diagram in Fig.2.
The function $F$ here is calculated as
\begin{eqnarray}
  &~&
  F(\omega_{n},\vec{p})^{-1}
  =
  -\,\omega_{n}^2 + \vec{p}\,^2 + M^2(T)
  + \int^\infty_{-\infty} d\omega\,
  \frac{r_-(\omega,\vec{p})}{\omega - \omega_n}
\,, \\
  &~&
  M^2(T)
  = M^2 +
  \frac{\lambda }{24}\,T^{2}
  \nonumber \\ 
  &~&
  -\, \frac{\lambda ^{2}}{32\pi ^{3}}(\frac{MT}{2\pi })^{1/2}e^{-M/T}\,
  \int_{0}^{\infty }\,dk\,\frac{k^{2}}{\omega _{k}^{2}}
  \left( \frac{e^{-\beta \omega _{k}}}{(1 - e^{-\beta \omega _{k}})^{2}}
  + \frac{1}{\beta \omega _{k}(e^{\beta \omega _{k}} - 1)}\right)
\,, \label{mass shift} 
\end{eqnarray}
valid at $m \ll T\ll M$ where $\omega _{k} = \sqrt{\vec{k}\,^{2} + m^{2}} $.
The temperature dependent mass $M^{2}(T)$ has correction both of
$\lambda $ and $\lambda ^{2}$ orders, but the $O[\lambda ^{2}]$ term
is Boltzmann suppressed.
In the function $F$ the proper self-energy given by a $r_{-}$ integral
is summed up to get
a geometric form, which is allowed to $O[\lambda ^{2}]$.
The discrete energy sum here may be converted by a contour integration
to the form given by the real-time correlator, which is nothing but 
eq.(\ref{asymp thermal correlator}) along with (\ref{overlap formula}).
Thus, two methods give the same result.

\vspace{1cm}
\section{Quantum kinetic equation
}

\hspace*{0.5cm} 
Derivation of the kinetic equation goes in a few steps.
We shall only sketch the main point of this sequence of arguments.
It starts from the self-consistency equation for the quantity
$\sigma (k \,, t)$, the Fourier transformed correlator.
We first note that the odd part of this quantity
\begin{equation}
\sigma _{-}(k \,, t) = \sigma (k \,, t) - \sigma (-k \,, t)
\end{equation}
slowly varies. Thus, to order of $\lambda ^{2}$ one can use the infinite
time limit;
\begin{equation}
\sigma _{-}(k \,, t) = \sigma _{-}(k\,, \infty ) +
O[\lambda ^{4}] \,.
\label{approximate anti-sigma} 
\end{equation}
The basic reason for this simplicity is due to a symmetry of
certain terms of eq.(\ref{correlator form})
under $x_{0} \leftrightarrow y_{0}$.

The even part, or a more convenient quantity 
\begin{equation}
\tau ( k \,, t) = \frac{\sigma (-k \,, t)}{\sigma _{-}(k\,, t)}
\end{equation}
which is constrained by 
\begin{equation}
\tau (k_{0}, \vec{k}\,, t) + \tau (-\,k_{0}\, \vec{k} \,, t) = -1 \,, 
\end{equation}
is used for derivation of the kinetic equation.
Under the condition (\ref{approximate anti-sigma}) 
one may use the "constant" 
$\sigma _{-}(k \,, \infty )$, which can be replaced by
\begin{equation}
2\pi H(k\,, \infty ) = \sigma _{-}(k \,, \infty ) \,.
\end{equation}

The kinetic equation is derived from the self-consistency relation
symbolically written as
\begin{equation}
\tau (t) = T[\tau (t) \,, t] \,, 
\end{equation}
where the explicit and implicit (via the function $\tau $) time dependence
is written. The right hand side is understood as a functional of $\tau (t)$.
By dropping higher order $O[\lambda ^{4}]$ terms, we may derive
from this equation
\begin{eqnarray}
&&
\frac{d\tau }{dt} = -\,\Gamma \,\left( \,
\tau - T[\tau (t) \,, \infty ]\,\right)
 \,,
\\ &&
\Gamma = -\,\frac{\partial }{\partial t}\,\ln \left( \,
T[\tau (t) \,, t] - T[\tau (t) \,, \infty ]\,\right) \,.
\end{eqnarray}
The neglected term is $\partial T/\partial t$ of order $\lambda ^{2}$
and it was replaced as
\begin{equation}
1 - \frac{\partial T}{\partial \tau } \:\rightarrow \; 1 \,,
\end{equation}
in the left hand side.
One explicitly works out this symbolic equation for our case and
in the end  derives the quantum kinetic equation in the form,
\begin{eqnarray}
 &~&\frac{d \tau(p,t)}{dt}
  =
  \frac{\lambda^2}{2p_0}
  \int \frac{d^3k_{1}}{(2\pi)^32\omega_{k_{1}}}
  \int \frac{d^3k_{2}}{(2\pi)^32\omega_{k_{2}}}
  \int \frac{d^3p'}{(2\pi)^3}
  \int^\infty_0 \frac{dp'_0}{2\pi}\,
  (2\pi)^4\sigma_-(p'\,, \infty )
  \nonumber \\
  &~&
  \left\{ \,
    \delta^{(4)}(p + p' - k_1 - k_2)
    \left[
      \tau \tau'(1 + f_1)(1 + f_2) - (1 + \tau )(1 + \tau')f_1f_2
    \right]
  \right.
  \nonumber \\
  &~& +
  2\delta^{(4)}(p + p' + k_1 - k_2)
  \left[
    \tau \tau'f_1(1 + f_2) - (1 + \tau )(1 + \tau')(1 + f_1)f_2
  \right]
  \nonumber \\
  &~& + 
  \delta^{(4)}(p + p' + k_1 + k_2)
  \left[
    \tau \tau'f_1f_2 - (1 + \tau )(1 + \tau')(1 + f_1)(1 + f_2)
  \right]
  \nonumber \\
  &~& +
  \delta^{(4)}(p - p' - k_1 - k_2)
  \left[
    \tau (1 + \tau')(1 + f_1)(1 +f_2) - (1 + \tau )\tau'f_1f_2
  \right]
  \nonumber \\
  &~&  +
  2\delta^{(4)}(p - p' + k_1 - k_2)
  \left[
    \tau (1 + \tau')f_1(1 + f_2) - (1 + \tau )\tau'(1 + f_1)f_2
  \right]
  \nonumber \\
  &~&
  \left. +
    \delta^{(4)}(p - p' + k_1 + k_2)
    \left[
      \tau (1 + \tau')f_1f_2 - (1 + \tau )\tau'(1 + f_1)(1 + f_2)
    \right]\,
  \right\} \,.
  \label{quantum kinetic eq} 
\end{eqnarray}
In the right hand side $\tau = \tau (p\,, t) \,, 
\tau ' = \tau (p' \,, t)$.
In the final step the Markovian approximation for the rate
was used; the exponential decay law for the quantity
$T[\tau (t) \,, t]$ was assumed with the pole approximation.

The kinetic equation thus derived has a structural resemblance to
the conventional Boltzmann equation. There are however a number of
differences. The most important one is that there is no mass shell
condition for the $\varphi $ particle; $p_{0}\,^{2} - \vec{p}\,^{2} \neq 
M^{2}$.
Accordingly the function $\tau $ is a function of the 4-momentum $p$.
Related to this is that even the processes not allowed by the energy-momentum
conservation on the mass shell all contribute to the collision term
in the right hand side, for example $1 \leftrightarrow 3$ process
such as $\varphi \leftrightarrow \varphi \chi \chi $.
Even if the equilibrium solution $\tau (p \,, \infty ) 
= 1/(e^{\beta p_{0}} - 1)$ discussed in the preceeding section
coincides with the familiar Bose-Einstein form at $p_{0} = 
\sqrt{\vec{p}\,^{2} + M^{2}}$, 
the $p_{0}$ variable here
is defined all the way from $-\infty $ to $\infty $.
These are important differences, although the structural resemblance
is impressive and suggests deeper understanding.

Once the kinetic equation is solved, one can compute physical
quantities using the function $\tau $.
For instance, the $\varphi$ energy density is calculated taking
the coincident limit of two-point correlators, with due consideration
of renormalization of composite operators.
When Fourier transformed, the correlator is given by
\begin{equation}
\sigma (p\,, t) = \sigma _{-}(p\,, t)\,\tau (-\,p\,, t)
= 2\pi \,H(p\,, t)\,\tau (-\,p\,, t) \,.
\end{equation}
Hence
\begin{eqnarray}
  \langle \cal{H}_{\varphi} \rangle
  &=&
  \langle
  \frac{1}{2}\dot{\varphi}^2
  + \frac{1}{2}(\nabla{\varphi})^2
  + \frac{M^2}{2}\varphi^2 - ({\rm counter \; terms})
  \rangle
  \nonumber \\
  &=&
  \frac{1}{2}\int\frac{d^4p}{(2\pi)^3}\,
  \{p_0^2 + E_p^2\}H(p\,, t)\tau(p,t)
  - ({\rm counter \; terms})
  \nonumber \\
  &=&
  \int\frac{d^3p}{(2\pi)^3}\int^\infty_0 dp_0\,
  \{p_0^2 + E_p^2\}H(p\,, t)\tau(p,t)
  \nonumber \\
  &+&
  \frac{1}{2}\int\frac{d^3p}{(2\pi)^3}\int^\infty_0 dp_0\,
  \{p_0^2 + E_p^2\}H(p\,, t)
  - ({\rm counter \; terms})
  \,.
\end{eqnarray}
This is further simplified using
$H(p\,, t) \approx H(p\,, \infty )$ and
\begin{eqnarray}
  &~&
  \frac{1}{2}\int\frac{d^3p}{(2\pi)^3}\int^\infty_0 dp_0\,
  \{p_0^2 + E_p^2\}H(p\,, \infty )
  \nonumber \\
  &=&
  \frac{1}{2}\int\frac{d^3p}{(2\pi)^3}\int^\infty_0 dp_0\,
  \frac{r_-(p\,, \infty )}{(p_0 + E_p)^2} + \frac{1}{2}
  + O[\lambda ^{3}]
  \,.
\end{eqnarray}
Thus,
\begin{eqnarray}
  \langle \cal{H}_{\varphi} \rangle
  &\simeq &
  \int\frac{d^3p}{(2\pi)^3}\int^\infty_0 dp_0\,
  \{p_0^2 + E_p^2\}H(p\,, \infty )\tau(p,t)
    + c\,\lambda^2\frac{T^6}{M^2}
    \nonumber \\
  &\simeq&
  \int\frac{d^3p}{(2\pi)^3}\,
  E_p\, \tau(E_p,\vec{p}\,,t)
  + c\,\lambda^2\frac{T^6}{M^2} \,, \label{energy density} 
  \\
  c &=& \frac{1}{69120} \sim 1.4\,\times 10^{-5} \,,
\end{eqnarray}
dropping Boltzmann suppressed $O[\lambda ^{2}]$ terms.
In the last step we replaced the spectral weight $H$ by the
delta function, since perturbative analysis works for this
quantity.
In this computation we assumed the mass hierarchy for the light
$\chi $ particle, $m \ll T$.
We used in this derivation the formula (\ref{r_ using f}) to get
($T \ll M$ assumed)
\begin{eqnarray}
  &~&
  \frac{1}{2}\int\frac{d^3p}{(2\pi)^3}\int^\infty_0 dp_0\,
  \frac{r_-(p\,, \infty )}{(p_0 + E_p)^2} + \frac{1}{2}
  - (\rm{counter~terms}) 
  \nonumber \\
  && \hspace*{-0.5cm} \simeq
   2\lambda ^{2}\,\int\,d\Pi _{k}\,\int\,d\Pi _{k'}\,
  (k^{2} + k'\,^{2})\,f_{k}f_{k'}\,\int\,d\Pi _{p}\,\int\,d\Pi _{p'}\,
  \frac{(2\pi )^{3}\delta ^{(3)}(\vec{p} + \vec{p}\,' + \vec{k}
  + \vec{k}\,')}{(E_{p} + E_{p'})^{3}}
  \nonumber \\
  && \hspace*{0.5cm} \sim
  \frac{\lambda ^{2}}{32\pi ^{2}}\,\int\,d\Pi _{k}\,\int\,d\Pi _{k'}\,
  (k^{2} + k'\,^{2})\,f_{k}f_{k'}\,\int_{0}^{\infty }\,dp\,
  \frac{p^{2}}{E_{p}^{5}}
  =
  c\,\lambda^2\frac{T^6}{M^2} \,, 
\end{eqnarray}
dropping Botzmann suppressed minor terms.
Physical processes that contribute to this equilibrium result are
\begin{equation}
\chi \chi \rightarrow \varphi \varphi ' \,, \hspace{0.5cm} 
\chi \rightarrow \varphi \varphi ' \chi \,, 
\end{equation}
$\varphi '$ being the partner $\varphi $ particle.
All other channels are suppressed by the Boltzmann factor,

The time dependence of physical quantities appears via the
function $\tau (E_{p} , \vec{p}\,, t)$
in eq.(\ref{energy density}).
This function defined on the mass shell ($E_{p} =
\sqrt{\vec{p}\,^{2} + M^{2}}$) follows the usual Boltzmann equation,
as seen from the kinetic equation (\ref{quantum kinetic eq}) along with
the approximation,
\begin{equation}
\sigma _{-}(p \,, \infty ) \approx \frac{2\pi }
{2E_{p}}\,\delta (p_{0} - E_{p}) \,.
\end{equation}
It is thus clear that in a state near thermal equilibrium
$\tau (E_p,\vec{p}\,,t) \sim 1/(e^{E_p/T} - 1)$.
Hence for $T/M \ll 1/\sqrt{c}\lambda \approx 3\times 10^{2}/\lambda $
\begin{eqnarray}
  \langle {\cal H}_{\varphi } \rangle
    \sim \langle {\cal H}_{\varphi } \rangle_{{\rm eq}} = 
  \int\,\frac{d^{3}p}{(2\pi )^{3}}\,\frac{\sqrt{p^{2} + M^{2}}}
  {e^{\sqrt{p^{2} + M^{2}}/T} - 1} +
  c\,\lambda ^{2}\frac{T^{6}}{M^{2}}
  \,.
\end{eqnarray}
The same equilibrium result was derived using the thermal field theory in
\cite{my-pair-99-2}.
Derivation given here is a separate and independent confirmation
of the equilibrium density from the kinetic approach.
The temperature power thus derived ($\propto T^{6}$) differs from that
given in ref.\cite{my-pair-99}, which should be corrected.

As discussed in \cite{my-pair-99-2}, the time evolution equation
in cosmology is approximately given by (at $m \ll T \ll M$)
\begin{eqnarray}
&&
\frac{dn_{\varphi }}{dt} + 3H\,n_{\varphi } = -\,
\langle \sigma v \rangle\,\left( n_{\varphi }^{2} - n_{{\rm eq}}^{2}\right)
 \,,
\\ &&
n_{{\rm eq}} \sim (\frac{MT}{2\pi })^{3/2}e^{-M/T} +
c\,\lambda ^{2}\frac{T^{6}}{M^{3}} \,,
\end{eqnarray}
where $H$ is the Hubble rate $\sim 1.7\sqrt{N}\,T^{2}/m_{{\rm pl}}$.
Numerical analysis shows that the picture of the sudden freeze-out
\cite{lee-weinberg} 
holds such that until the freeze-out the number density follows
the equilibrium value $n_{{\rm eq}}$.
Thus, the most important part of cosmological application is
to estimate the freeze-out temperature using the new
equilibrium number density.
This estimate is done by equating the annihilation rate
$\langle \sigma v \rangle\,n_{{\rm eq}}$ to the Hubble rate $H$.

The close connection with the equilibrium thermal field theory
is made evident by taking  the infinite central time limit;
the non-equilibirum correlator defined here approaches the equilibrium
value given by the thermal field theory,
\begin{eqnarray}
\langle \varphi (x)\varphi (y) \rangle =
{\rm tr}\;\rho _{i}\,\varphi (x)\varphi (y)
\;\rightarrow \;
{\rm tr}\;e^{-\beta \,H_{{\rm tot}}}\,\varphi (x)\varphi (y)/
{\rm tr}\;e^{-\beta \,H_{{\rm tot}}}
 \,.
\end{eqnarray}
The equilibrium correlator is time-translation invariant, hence
the central time dependence disappears for the correlator
and the non-equilibrium correlator recovers this invariance.
Moreover, our formula for the correlator and physical quantities such as
the energy density indicates that the approach towards 
thermal equilibrium described by the first term in eq.(\ref{energy density})
is essentially determined by a thermally averaged
rate of on-shell quantities, consistent with the previous
estimate of this quantity in derivation of the kinetic
equation.

Thus, the most important change for cosmological estimate of
the relic mass density is the new equilibrium quantity.
This equilibrium value is precisely given by the Gibbs formula,
using the total hamiltonian, not its free part.
This is the main reason the use of thermal field theory
in our companion paper \cite{my-pair-99-2} is justified in computation of
the relic mass density.
It has to be kept in mind that
the simple ideal gas form of distribution function must be
questioned in the low temperature region.
Even a higher order term in coupling is more important than
the zero-th order Boltzmann suppressed term 
when the Boltzmann suppression factor is huge.
The appearance of our temperature power term for the equilibrium
number density is due to
a continuous integral of these Boltzmann suppressed terms
which dominates over a single contribution of quasi-particle
object given by approximating a Breit-Wigner form of integral
by a single complex pole.

Despite this rather obvious change, the thermally
averaged Boltzmann equation is hard to justify at low and intermediate
temperatures, or
its slight modification is difficult to lead to the relevant correction.
Although our work justifies in some sense the Boltzmann equation
at high temperatures, it should be kept in mind that
full quantum mechanical treatment is essential to derive
the correct low temperature result.

\vspace{1cm}
\begin{center}
{\bf Acknowledgment}
\end{center}

The work of Sh. Matsumoto is partially
supported by the Japan Society of the Promotion of Science.

\newpage
\begin{Large}
\begin{center}
{\bf Figure caption}
\end{center}
\end{Large}

\vspace{0.5cm} 
\hspace*{-0.5cm}
{\bf Fig.1}

Diagram for the spectral weight of two light $\chi $ particle
intermediate states denoted by broken lines.

\vspace{0.5cm} 
\hspace*{-0.5cm}
{\bf Fig.2}

Diagram for thermal correlator to $O[\lambda ^{2}]$.
The solid and broken lines correspond to the $\varphi $ and $\chi $
particle propagators in thermal equilibrium, respectively.

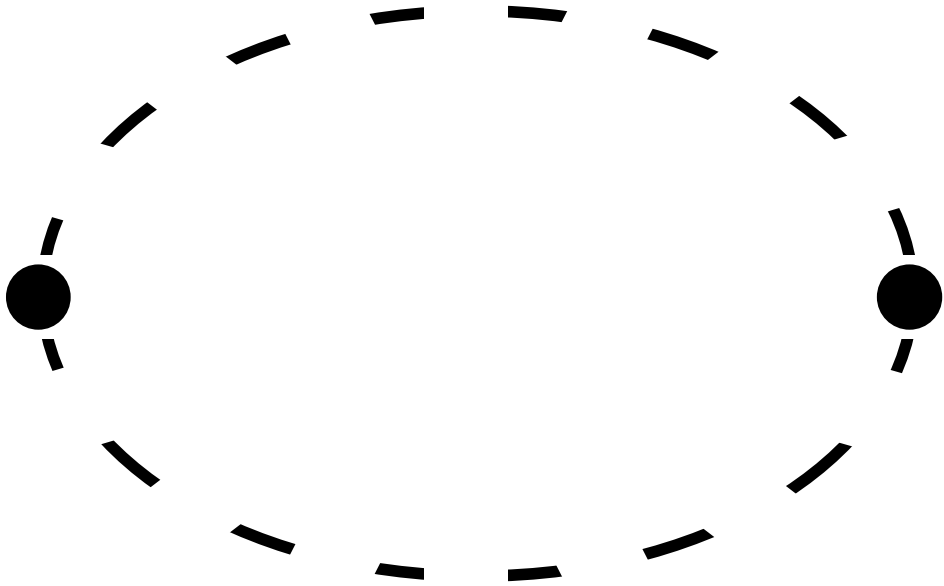
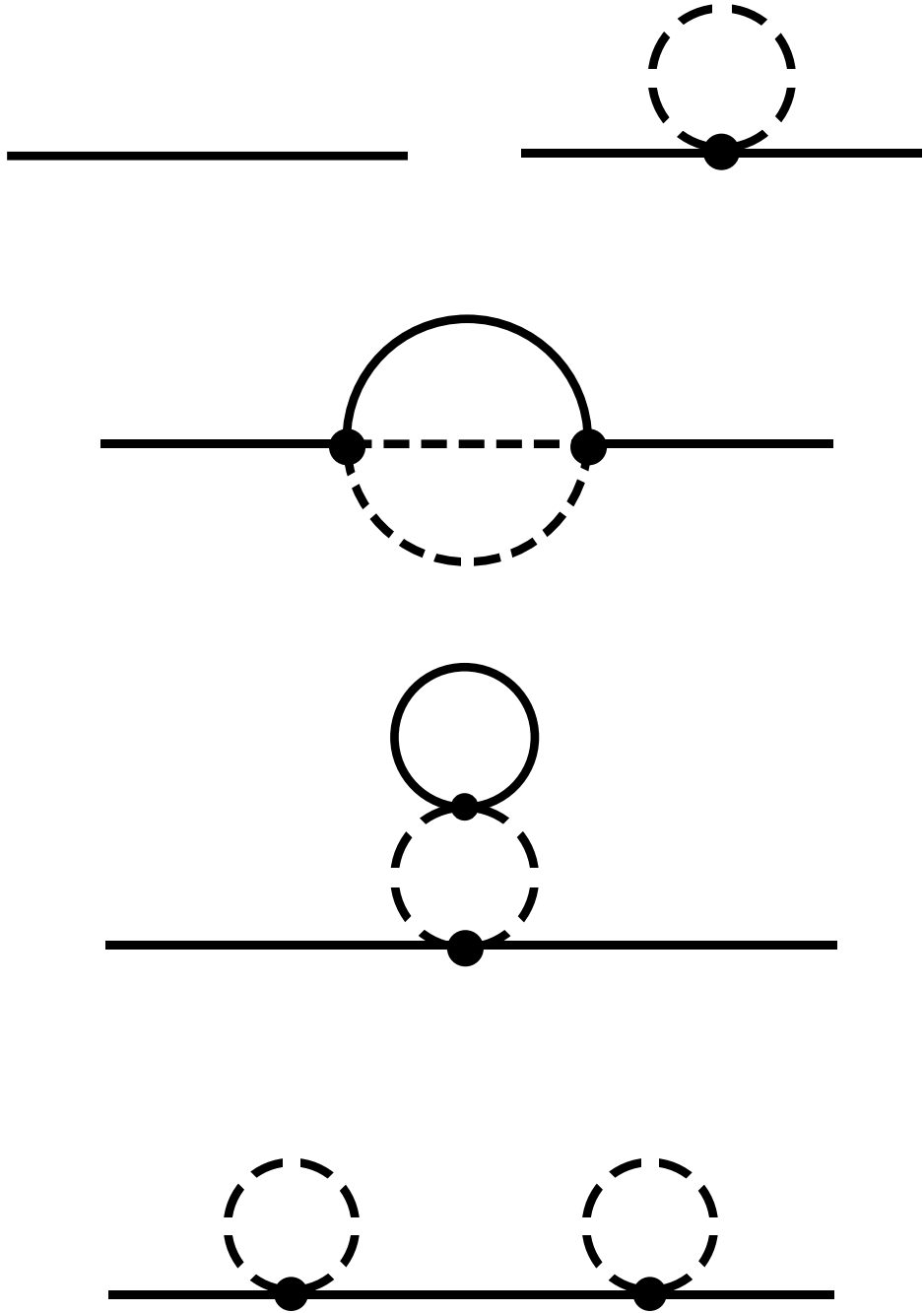
\end{document}